\documentclass[reprint,prl,longbibliography,twocolumn,superscriptaddress,english,showpacs]{revtex4-1} 
\usepackage[english]{babel}
\usepackage[dvipsnames]{xcolor} 
\usepackage{graphicx} 
\usepackage{amsmath} 
\usepackage{amssymb}

\definecolor{darkblue}{rgb}{0.1,0.2,0.6} \definecolor{darkred}{rgb}{0.8,0.1,0.2}
\usepackage[colorlinks,citecolor=darkblue,linkcolor=darkred,urlcolor=darkblue]{hyperref}
\usepackage[all]{hypcap} 

\newcommand{\ket}[1]{|#1\rangle}

\newcommand{\mat}[1]{\mathsf{#1}}

\newcommand{\D}{\mathrm{d}}

\newcommand{\cf}{\textit{cf.} }

\usepackage{ifdraft}

\begin{document}
\title{{How a small quantum bath can thermalize long localized chains}} 
\author{David J. Luitz}
\affiliation{Department of Physics T42, Technische Universit\"at M\"unchen, James-Franck-Stra\ss e 1, 85748 Garching, Germany}
\email{david.luitz@tum.de}

\author{Fran\c{c}ois Huveneers}
\affiliation{Universit\'e Paris-Dauphine, PSL Research University, CNRS, CEREMADE, 75016 Paris, France}
\email{fhuvenee@ceremade.dauphine.fr}

\author{Wojciech De Roeck}
\affiliation{Instituut voor Theoretische Fysica, KU Leuven, Belgium}
\email{{wojciech.deroeck@kuleuven.be}}

\date{May 30, 2017}

\begin{abstract} 
We investigate the stability of the many-body localized (MBL) phase for a system in contact with a
single ergodic grain, modelling a Griffiths region with low disorder. Our numerical analysis
provides evidence that even a small ergodic grain consisting of only 3 qubits can delocalize a
localized chain, as soon as the localization length exceeds a critical value separating localized
and extended regimes of the whole system. We present a simple theory, consistent with the arguments
in [Phys. Rev. B \textbf{95}, 155129 (2017)], that assumes a system to be
locally ergodic unless the local relaxation time, determined by Fermi's Golden Rule, is larger than
the inverse level spacing. This theory predicts a critical value for the localization length that is
perfectly consistent with our numerical calculations.  We analyze in detail the behavior of local
operators inside and outside the ergodic grain, and find excellent agreement of numerics and theory.
\end{abstract} 

\maketitle

\textbf{Introduction} --- 
The phenomenon of Many-Body Localization (MBL)
\cite{anderson_absence_1958,fleishman_interactions_1980,basko2006metal,gornyi2005interacting,znidaric_many-body_2008,oganesyan_localization_2007,pal_many-body_2010,luitz_many-body_2015,nandkishore_many-body_2015,altman_universal_2015,abanin_recent_2017,luitz_ergodic_2017,agarwal_rare-region_2017,parameswaran_eigenstate_2017,imbrie_review:_2016} has
challenged our ideas on thermalization and the applicability of thermodynamics. It is hence
important to determine the precise conditions for the stability of the MBL phase.  Whereas  in
the original works \cite{basko2006metal,gornyi2005interacting}, the spatial dimension $d$ did not
play a central role, the rigorous treatment of Griffiths regions of low disorder\footnote{In contrast, Griffiths regions of high disorder are also believed to be responsible for anomalous
    transport prior to the MBL transition
    \cite{bar_lev_dynamics_2014,bar_lev_absence_2015,agarwal_anomalous_2015,luitz_extended_2016,znidaric_diffusive_2016,varma_energy_2017,agarwal_rare-region_2017,luitz_ergodic_2017}. } in
\cite{imbrie2016many} relies on $d=1$, not for technical but for conceptual reasons.  More
generally, it is now well understood that there is a huge variety of systems where thermalization is
effectively inhibited locally and only rare Griffiths regions can, possibly, restore ergodicity 
\cite{huveneers2017classical,de2014scenario}. This applies in particular to quasi-localization
\cite{de2014asymptotic,pino2016nonergodic,grover2014quantum,yao2014quasi,schiulaz2013ideal,schiulaz2015dynamics, Nandkishore_Sondhi_2017},
classical disordered models \cite{huveneers2013drastic,basko2011weak} or even glasses
\cite{hickey2016signatures,berthier2011theoretical}.  It is also believed that Griffiths regions
drive the transition from MBL to ergodicity
\cite{vosk2015theory,potter_universal_2015,pekker_hilbert-glass_2014,dumitrescu2017scaling,thimothee,khemani2017critical}. This
raises the fundamental issue of understanding the effect of a Griffiths region, in practice an ergodic
grain or \emph{imperfect bath}, on a localized system. We outline a very simple theory: local ergodicity characterized by the
Eigenstate Thermalization Hypothesis (ETH)
\cite{feingold1984ergodicity,deutsch1991quantum,srednicki1994chaos,rigol_thermalization_2008,khatami_quantum_2012,d2015quantum,borgonovi_quantum_2016} is taken as the default option, and a
degree of freedom is interpreted as localized if the local relaxation time would be larger than the
inverse level spacing. In \cite{de_roeck_stability_2016} a more microscopically motivated version of
this theory was proposed leading to precisely the same conclusions. The main result is the instability of MBL
if the bare localization length is larger than a critical value %
\footnote{A bound on the
    localization length has also been derived in \cite{vosk2015theory}, which is entirely different
    (it holds even without ergodic regions) and for a different quantity (the typical matrix element
    of the coupling whereas the bound in \cite{de_roeck_stability_2016} is for the norm of the coupling). The crucial difference
    between these bounds is discussed in \cite{thimothee} }.  
This predicted instability implies that a single, sufficiently large, interacting, ergodic
grain thermalizes the whole system \textit{if the localization length in the localized part of the system is large enough}. 
This striking conclusion is counter-intuitive, and it has often been suggested to us, e.g.\ \cite{altmanprivate}, that localization should prevail when the number of a priori localized degrees of freedom clearly exceeds the number of degrees of freedom in the bath.  
 In this letter, we investigate
this aspect, using a setup where the ergodic grain is considerably smaller than the surrounding localized
system, namely 3 versus 13 spins. All results of our numerical analysis confirm the simple
theory, leading to thermalization of the chain by a small ergodic grain.

\begin{figure}[t]
    \centering
    \includegraphics[draft=false,width=\columnwidth]{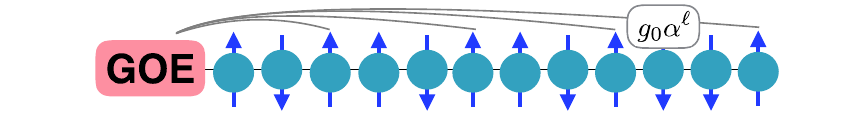}
    \caption{LIOMs (blue) coupled to an ergodic grain (red) modelled by a GOE matrix.}
    \label{fig:BathLIOMs}
\end{figure}

\textbf{Theoretical predictions} --- Let us consider a system of size $L=L_\text{loc}+L_b$
containing an ergodic grain (or bath) of size $L_b$  interacting with a fully localized
chain of size $L_\text{loc}$. The full system is described by the Hamiltonian  $H = H_b + H_{\text{loc}} + H_{bl}$.  We assume that the bath
Hamiltonian $H_b$ satisfies the \textit{eigenstate thermalization hypothesis} (ETH). $H_{\text{loc}}$
describes an Anderson insulator in the basis of localized integrals of motion (LIOMs)
\cite{chiara_entanglement_2006,znidaric_many-body_2008,bardarson_unbounded_2012,serbyn_universal_2013,serbyn_local_2013,huse_phenomenology_2014,ros_integrals_2015,imbrie_review:_2016, rademaker2017many,monthus_many-body_2016}, and 
$H_{bl}$ couples the bath locally to the LIOMs. A LIOM at distance $\ell$ from the bath
is connected to the bath by a coupling strength $g_{\ell}:=g_0 e^{-\ell/\xi} = g_0 \alpha^\ell$ reminiscent of
the interaction of an MBL system with the grain, where $\alpha
= e^{-1/\xi}$ and where $\xi$ denotes the localization length (in lattice units); \cf
Fig. \ref{fig:BathLIOMs} and Eq.~\eqref{eq:Hnosym} for the model used in the numerics. 

ETH provides an ansatz for the matrix elements of a local operator $A$ in an ergodic system. 
Let $E,E'$ label eigenstates with energy density $\epsilon_{0}= E/L\sim E'/L$, and let $\langle A\rangle_{\epsilon_{0}}=0$ (ensemble average in an equilibrium state at $\epsilon_{0}$). 
Below we always let $\epsilon_0$ correspond to maximal entropy and we drop $\epsilon_0$ from the notation. 
Then, ETH amounts to \cite{d2015quantum}
\begin{equation}
    \langle E' |A |E \rangle  = \sqrt{\delta} f(\omega) \eta_{E,E'} \quad (E \ne E')
	\label{eq:ETHansatz}
\end{equation} 
with $\delta$ the many-body level spacing at density $\epsilon_{0}$,  $\omega=E-E'$ the energy
difference, $\eta_{E,E'}$ random numbers with zero mean and unit variance and $f(\omega)$ a smooth function that can be
related to a time-dependent correlation function and that satisfies the sum rule $\int d\omega |f(\omega)|^2 =\langle A^\dagger A \rangle_{\epsilon_{0}} \sim 1$.
For local $A$, $f(\omega)$ is roughly supported on intervals of size $ \Delta$, where $\Delta$
can be interpreted as the typical rate at which local quantities
equilibrate \cite{luitz_anomalous_2016,serbyn_thouless_2016,d2015quantum}
\footnote{In well coupled ergodic systems where all local terms are more or less of equal strength, one can think of $\Delta$ as roughly equal to $\epsilon_0$, i.e.\ the energy per site.  There is however a glitch in calling this the 'local rate': In ergodic systems with conserved quantities and normal (diffusive) transport, the relaxation time of conserved quantities scales as $L^2$ and the associated rate is the Thouless energy. A generic local observable will inherit this slow decay and so its relaxation rate, strictly speaking, is also given by the Thouless energy. Yet, there is a seperation of time scales and for short times, the rate appears to be $\Delta$.   In any case, these distinctions do not matter in our work since we focus on the exponential dependence on $L$.
}. 
Obviously, the ansatz \eqref{eq:ETHansatz}  only makes sense
if $\Delta \gg \delta$ (otherwise \eqref{eq:ETHansatz} conveys no information), which we use as a \emph{consistency condition} that will be invoked below in a crucial way.
Let us give a relevant example of how to determine $\Delta$: take one LIOM with field $h$ coupled to an ergodic system (e.g.\ our ergodic grain) via a weak coupling term of strength $g$.  
Then, under some mild conditions, Fermi's Golden Rule predicts $f(\omega)$  of that LIOM in the combined system to have
principal peaks at the Bohr frequencies $\pm 2 h$  with widths of order $\Delta=g^2/\epsilon$, for some local energy scale $\epsilon$ characterizing the bath, see e.g.\@ \cite{de_roeck_stability_2016}.   

Now, we couple LIOMs $i=1,\ldots, L_{\text{loc}}$ to the
bath (\cf Fig. \ref{fig:BathLIOMs}) with couplings $g_i=g_0 \alpha^{i}$ and compute the local rates
$\Delta_i$ as in the example above, namely $\Delta_i \sim \alpha^{2i}g^2_0/\epsilon$. Now we assume
that \emph{ETH is valid as long as it is not inconsistent}, i.e.\ as long as $ \delta \ll \Delta_i $
for all $i \leq L_\text{loc}$.  This central assumption will lead to striking consequences, presented below.  Verifying these consequences is the main point of the paper. 
We conjecture hence that ETH for the full system is valid if and only if
\begin{equation} \label{eq: thermalization condition}
    (g_0^2/\epsilon) \alpha^{2L_\text{loc}}\geq \mathcal{W} 2^{-L},
\end{equation}
where we put $  \delta= \mathcal{W} 2^{-L}$ with $\mathcal{W}=\epsilon_0 L$ the spectral width. Neglecting all non-exponential dependence on the lengths $L_{\text{loc}},L_b$, the condition for full ETH becomes (leading order in $L_b\gg 1$)
\begin{equation}
    L_\text{loc} \leq  L_c  = \frac{L_b \log 2}{\log\alpha^{-2} -  \log 2} 
	\label{eq:InstabilityCondition_d1}
\end{equation}
The most striking consequence of this analysis is that $L_c=\infty$ if $\alpha>\alpha_c=\frac{1}{\sqrt 2}$ (and $L_b$ is not too small). That is: a small bath is capable of thermalizing an
arbitrary number of LIOMs. 
This is a sharp prediction that can be tested numerically by studying the validity of ETH as a
function of $\alpha$.  

Let us add a more quantitative prediction of the theory: The local rates
$\Delta_i$ correspond to an \emph{effective dimension} $d_i= \Delta_i/\delta$, i.e.\ the number of
eigenstates $|E'\rangle$ over which $A_i|E\rangle$ is spread out, according to \eqref{eq:ETHansatz},
for a local operator $A_i$ at site $i$.  Then our theory predicts that 
\begin{equation} \label{eq: effective dimension}
d_i\approx 2^{L_{\text{therm}}} \alpha^{2i},\qquad L_{\text{therm}}=L_b+\min (L_{\text{loc}},L_c),
\end{equation}
as long as $d_i\geq 1$, and setting $d_i= 1$ otherwise, corresponding to localization at site $i$. 
Here, $L_{\text{therm}}$ is the length of the thermal region and for $\alpha>\alpha_c$, we simply
have $L_{\text{therm}}=L$.  
In the rest of this letter, we present the several numerical results that demonstrate the accuracy
of the above picture for numerically accessible system sizes.

\textbf{Model} --- 
\begin{figure}[t]
    \centering
    \includegraphics[draft=false]{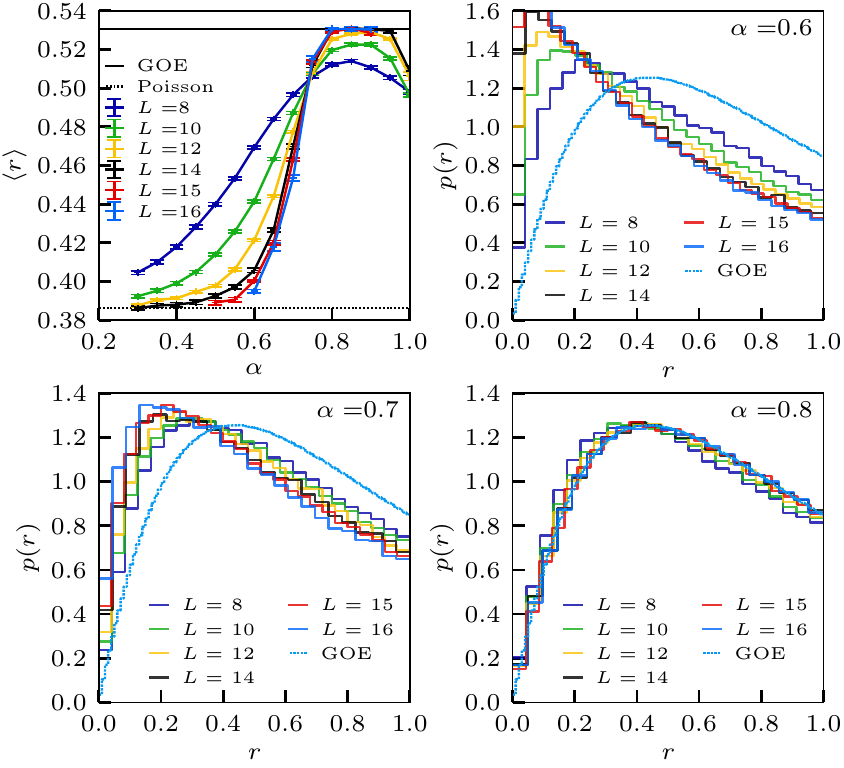}
    \caption{Probability distributions and average of the adjacent gap ratio $r$ for different coupling strengths
    and system sizes compared to the distribution obtained for large GOE matrices. }
    \label{fig:gaphisto}
\end{figure}
We start from a simplified model of an ergodic grain coupled to localized spins, as depicted on
Fig. \ref{fig:BathLIOMs}, where the Hamiltonian of the ergodic grain is given by a random matrix
$\mat R$ drawn from the GOE ensemble: $\mat{R} = \frac{\beta}{2} \left( \mat{A} + \mat{A}^T \right) \in
    	\mathbb{R}^{2^{L_b}\times 2^{L_b}}$ where 
$\mat{A}_{ij} = \text{norm}(0,1)$ with
norm(0,1) random numbers drawn from a normal distribution with zero mean and unit
variance.

Now, we can specify the Hamiltonian for our model for an ergodic grain of size $L_b$ and a
set of LIOM's on sites $0,\dots ,L_{\text{loc}}-1$. 
\begin{equation}
    H = \mat R + \sum_{i=0}^{L_{\text{loc}}-1} \frac{h_i}{2} \sigma_i^z + 
    \sum_{i=0}^{L_{\text{loc}}-1} \frac{g_0 \alpha^i }{4} \sigma_i^x \sigma_{-1}^x.
    \label{eq:Hnosym}
\end{equation}
The system has no conservation
laws except energy and the total Hilbert space dimension is $\text{dim}(\mathcal H)=2^L$. We restrict the size of the
ergodic grain to $L_b=3$ (on sites $-3,-2,-1$, hence $\sigma_{-1}^x$ is a bath operator).
For a more accurate correspondence of the LIOMs to a generic MBL system, we would need to include
interactions of the type $\sigma_i^z\sigma^z_{i+1}$. For simplicity, we omit those interactions,
thus making our localized chain basically an Anderson Insulator (AI). This simplification does not change the physics, since anyhow the bath coupling makes
the full system truly interacting. 
Our theory is hence not in conflict with the fact that one can construct AIs with arbitrarily large
localization lengths. 

The onsite fields $h_i$ are drawn from a random box distribution $[1-W,1+W]$,
$W=0.5$.  The shift of the box distribution by 1 is not necessary but reduces finite size effects.
We have checked carefully that a symmetric distribution around 0 yields similar results.  
In all our experiments, $L_{b} = 3$, $\beta = 0.3$, $g_0 = 1$. We have selected these values so as to obtain the cleanest results with the
smallest bath size and the smallest coupling constant $g_0$, however our results remain
qualitatively similar if any of these three parameters is varied (with $L_{b} \ge 3$).  In order to
infer the behavior of the system in the thermodynamic limit, we vary the number $L_{\text{loc}}$ of
localized spins coupled to the bath, with $L_{\text{loc}} \le 13$.  For $L_{\text{loc}}=13$, the
coupling strength of the last spin is $g_\text{min} = g_0 \alpha^{12}$.  For $\alpha = 0.8 >
\alpha_c$, the direct coupling to the bath is $0.8^{12}\approx0.068$ and does not suffice to
trivially thermalize the last few spins (\cf Supplemental Material). Thus, for such values of $\alpha$, the
thermalization of the last spins results from highly non-trivial effects involving all spins in between.

\textbf{Spectral statistics} --- 
A powerful and very general measure of ergodicity of quantum systems are the statistical properties
of its energy spectrum, typically studied in the center of the spectrum.  The gap
ratio\cite{oganesyan_localization_2007,pal_many-body_2010} $r = \min
\{\Delta E_k, \Delta E_{k+1} \} / \max \{\Delta E_{k}, \Delta E_{k+1} \}$ with $\Delta E_k =
E_{k+1} - E_k$ is Poisson distributed for localized systems and GOE distributed for ergodic ones, 
as an account of level repulsion \cite{Oganesyan_Huse_2007}. In Fig.
\ref{fig:gaphisto}, we show the average of $r$ as a function of $\alpha$ for various system sizes
$L$. A clear tendency towards GOE statistics for $\alpha \gtrsim 0.7$  and towards
Poisson statistics for $\alpha \lesssim 0.7$ is visible, consistently with the theoretical value for
$\alpha_c\approx 0.7071$.  It is
important to note that in the supercritical regime, the increase of $\langle r \rangle$ with $L$ is
due to the addition of \emph{increasingly weakly coupled spins}. To get a more detailed picture
near the transition, we show the full distribution of $r$ for $\alpha = 0.6, 0.7,0.8$: For $\alpha =
0.6,0.7$ we observe a clear drift towards Poisson statistics as the system sizes increases, while no
deviation from the GOE statistics is observed at $\alpha = 0.8$ up to $L=16$ \footnote{The case
    $\alpha = 1$ is particular because the energy interaction between bath and LIOMs becomes much
    bigger than the bath energy as $L_{loc}$ grows large.  This induces effects
    \cite{Huse_Nandkishore_et_al_2015,Nandkishore_Gopalakrishnan_2016} that are not covered by the
theory developed here}.

\textbf{Local magnetizations} ---
\begin{figure}[t]
    \centering
    \includegraphics[draft=false]{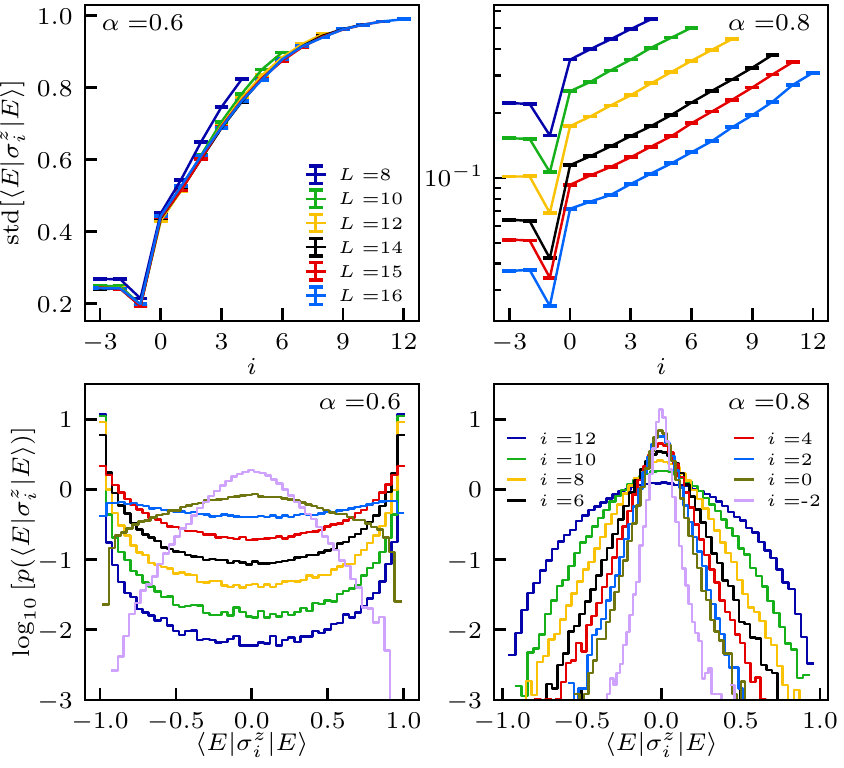}
    \caption{
    Top: Average over disorder realizations of the standard deviation of the eigenvector expectation
    of the magnetization at site $i$ ($i<0$ is in the bath, and $i\ge 0$ are LIOMs) for various
    total length $L$.
Bottom: Distribution of eigenvector expectation values of local magnetizations for $L=16$ for different positions in the chain.}
    \label{fig:maghisto}
\end{figure}
A direct test of the validity of ETH is furnished by the diagonal values of a local operator (here
$\sigma_i^z$) in the eigenbasis of the Hamiltonian: $\langle E |\sigma_i^z |E \rangle$. In a
localized system, the distribution of $\langle E | \sigma_i^z | E \rangle$ is sharply bimodal with
peaks at  $\pm 1$, since the LIOMs are small perturbations
of the bare spins $\sigma_i^z$. 
For an ergodic system, the distribution is sharply peaked around the thermal value (here: $0$), with variance scaling \cite{beugeling2014finite,ikeda2013finite} as $1/d_i$ with $d_i$ the
effective dimension, as computed in Eq. \eqref{eq: effective dimension}.
To test this, we show in Fig.~\ref{fig:maghisto} (top) the standard deviation over disorder and all eigenstates
$\ket{E}$ in a small window at maximal entropy of the expectation values $\langle E |\sigma_i^z |E
\rangle$, as a function of site index $-3 \le i < L_{\text{loc}}$
for $\alpha = 0.6 < \alpha_c$ and $\alpha = 0.8 > \alpha_c$.  At $\alpha =
0.6$, we observe that the standard deviation goes down slightly for $i< 0$ as we start increasing
$L$ but saturates quickly to a constant value.  This indicates that the first spins near the bath
are thermalized and increase the effective dimension for local operators in the bath, while spins further away, at $i>L_c$, remain localized and do not affect the effective dimension.
For $i\ge 0$, the standard deviation tends to its
maximal value 1 as the operator moves away from the bath, consistently with the fact that only the closest
spins get thermalized. The very good data collapse at large $L$ indicates that 
stationary values have been reached. The situation is strikingly different at $\alpha = 0.8$: at any
fixed distance $i$ from the bath, the
standard deviation decreases as the total length $L$ increases, because all spins contribute to
the effective dimension. On the other hand, for fixed $L$, the standard deviation always increases
as one moves away from the bath (i.e.\@ as $i$ increases). This is fully consistent with the decreasing effective dimension predicted in
\eqref{eq: effective dimension} and it should hence not be interpreted as
some sort of ``imperfect thermalization".  Finally, it is instructive to compare the standard deviation of
the last spin at different $L$, i.e.\ the endpoints of the curves: these endpoints move down with
increasing $L$. Thus, the last spins become more and more delocalized, \emph{even though they also move
further away from the original bath}.

This picture is even clearer in the full distribution of $\langle E |\sigma_i^z
| E \rangle$ shwon for $L=16$ for a subset of sites $i$ and $\alpha = 0.6,0.8$ in Fig.~\ref{fig:maghisto}
(bottom). 
At $\alpha = 0.6$, the distribution of all but the first few spins become strongly bi-modal with
peaks as $\pm 1$, indicating that spins far away form the bath are indeed not thermalized. In
contrast, for $\alpha = 0.8$, we see a progressive broadening of the distribution as $i\ge 0$ increases
but no signs of bi-modality, confirming thus the above conclusion. 
Interestingly, we observe that for all thermalized spins, the distribution departs from a gaussian
due to the presence of heavier tails compared to gaussian distributions. This phenomenon has
recently been observed in ergodic systems at moderate values of disorder where the dynamics is
expected to be sub-diffusive \cite{luitz_long_2016,luitz_anomalous_2016}.  

\textbf{Participation entropy} ---
\begin{figure}[t]
    \centering
    \includegraphics[draft=false]{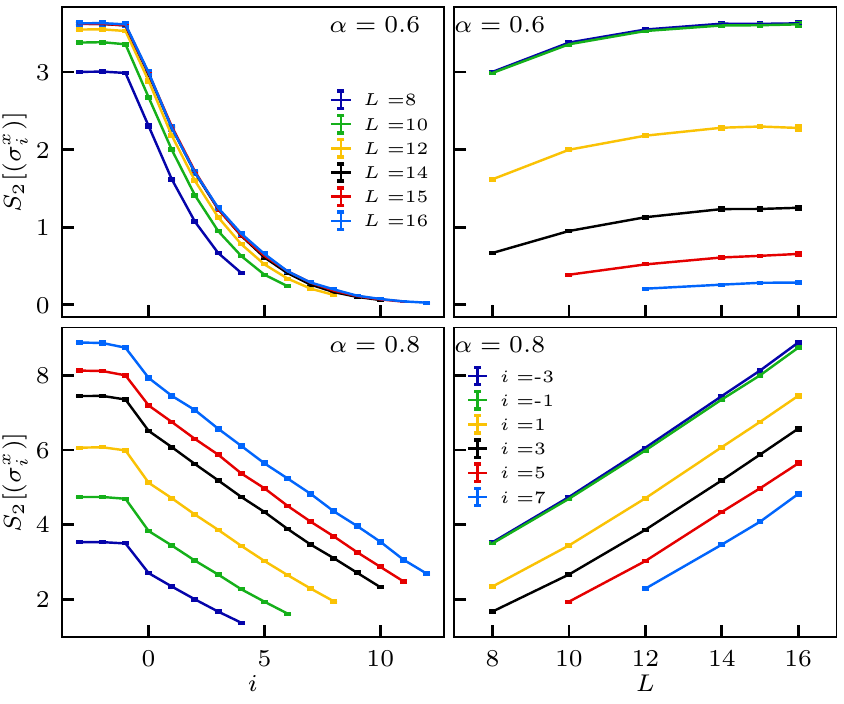}
    \caption{Left panels: Second R\'enyi entropy $S_2$  for excitation created by the operator $\sigma_i^x$ in an eigenstate. 
    Right panels: evolution of $S_2$ at some given sites as a function of the total length $L$.}
    \label{fig:logipr}
\end{figure}
Up to here, we have defined the ``effective dimension'' $d_i$ of the Hilbert space only via the spectral function $f(\omega)$. 
Let us introduce now the \emph{participation R\'enyi entropies} $S_q$ to obtain a more direct definition:
\begin{equation}
    S_q = \frac{1}{1-q} \log \sum_E |\langle E_0 | \sigma_i^x |E \rangle|^{2q}.
    \label{eq:participation_entropy}
\end{equation}
with $E_0$ a state of energy density $\epsilon_0$ (at maximal entropy).  Here, we will focus on the
second R\'enyi entropy $S_2$. Before taking the logarithm, this quantity is analogous to the inverse
participation ratio in one-particle localization\footnote{Therefore it was called IPR in
\citep{serbyn_thouless_2016,de_roeck_stability_2016}.}. For thermal systems at maximal entropy, $S_2
\sim L \log 2$ while for localized systems $S_2 \sim \mathrm{const.}$ as a function of the total
size $L$.  In general, using the ETH \eqref{eq:ETHansatz}, we find $S_2=  -\log(\delta \int \D\omega
|f(\omega)|^4) $. If now $f$ is mainly supported  on a set of size $\Delta_i$ and we use the sum-rule
$\int \D\omega |f(\omega)|^2 \sim 1$ (independent of $L$), then we find $S_2 \approx
\log(\Delta_i/\delta)$, which indeed equals $\log d_i$ by our definition of $d_i$.  By comparison
with \eqref{eq: effective dimension} the theory predicts  $S_2\sim  \log(2)L_{\text{therm}}-2|\log(\alpha)| i,$
(as long as $S_2\geq 0$). 

We test these predictions for $\alpha = 0.6 < \alpha_c$ and $\alpha = 0.8 >
\alpha_c$ in Fig.~\ref{fig:logipr}. Our results confirm the behavior that was already commented in
the section on local magnetizations. In particular, at $\alpha = 0.6$, see Fig.~\ref{fig:logipr} (top), we observe 
that $S_2$ first decreases linearly for small $i\ge 0$ then saturates to a constant
value, because $L_{\text{therm}}$ saturates to $L_c$.  The behavior of $S_2$ for a given site is plotted as a function of the total length in the
right panels of Fig.~\ref{fig:logipr}: $S_2$ saturates to a constant value at all sites.  
For $\alpha = 0.8$ instead, $S_2$
increases linearly with system size at all sites $i$, and decreases linearly with $i$ for a given
size $L$; since the decrease with $i$ is slower than the increase with system size, all spins can be
thermalized for arbitrary large system sizes.  The evolution of $S_2$ at a
given site as a function of $L$ shows a linear increase with a slope close to $\log (2)$ as
predicted theoretically. 

\textbf{Entanglement entropy} ---
\begin{figure}[t]
    \centering
    \includegraphics[draft=false]{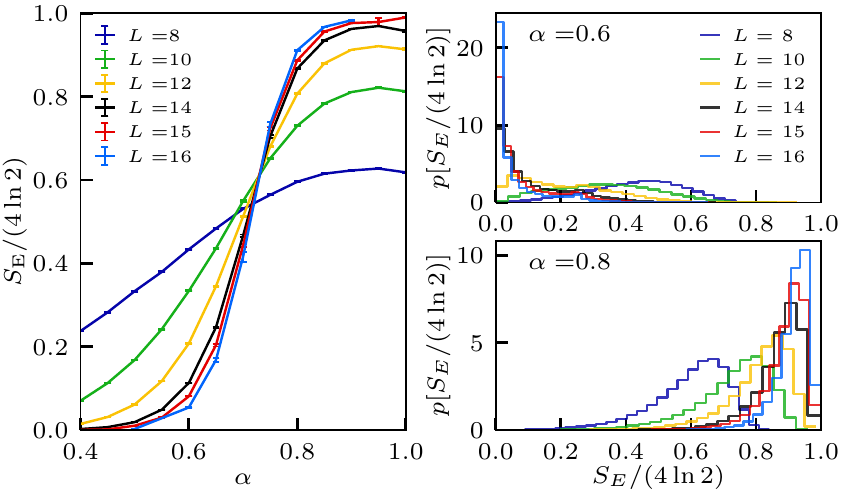}
    \caption{Entanglement entropy of the last 4 l-bits in the chain for different system sizes}
    \label{fig:EE}
\end{figure}
Finally, to have a more direct measurement of the thermalization of the last spins, we compute the
entanglement entropy (EE) of the last four spins together for an eigenstate in the middle of the
spectrum.  If the last four LIOMs remain localized the EE should remain close to $0$, and should
converge to $0$ as $L$ is increased.  Instead, if they are thermalized, the EE should approaches its
maximal value $4 \log (2)$.  The results are depicted on Fig.~\ref{fig:EE}.  In the left panel,
we clearly see the trend that, for $\alpha \lesssim 0.7$, the the value of the EE decreases to 0,
while it increases to $4 \log(2)$ for $\alpha \gtrsim 0.7$.  This is all the more remarkable since the 4 last
l-bits are increasingly far away from the bath with increasing system size. A full distribution is plotted
on the right panel for $\alpha = 0.6,0.8$ at various lengths, confirming this trend.

\textbf{Conclusion} ---  We have demonstrated that a small ergodic grain can thermalize an arbitrary
number $L_{\text{loc}}$ of localized spins, provided the localization length of the localized spins exceeds
the critical value $(2/\log 2)$. This was achieved numerically by coupling spins with on-site
disorder to a GOE system of dimension $8=2^3$ with exponentially decreasing couplings. 
When the localization length is smaller than the
critical value, the system drifts towards localization as more and more spins are added. When the
localization is above the critical value, the system drifts towards ever cleaner ergodicity as more
and more spins are added, even though those spins are coupled very weakly and can not be trivially
thermalized.

\begin{acknowledgments} 
    We would like to thank E. Altman, A. Chandran and J. Imbrie for useful discussions.
    This project has received funding from the European Union's Horizon 2020
    research and innovation programme under the Marie Sk\l{}odowska-Curie grant agreement No 747914
    (QMBDyn).  DJL acknowledges PRACE for awarding access to HLRS's Hazel Hen computer based in
    Stuttgart, Germany under grant number 2016153659.  FH benefited from the support of the projects
    EDNHS ANR-14-CE25-0011 and LSD ANR-15-CE40-0020-01 of the French National Research Agency (ANR).  WDR acknowledges the support of the Flemish Research Fund FWO and 
\end{acknowledgments}

\appendix
\section*{Supplementary Material}

\begin{figure}[h]
    \centering
     \includegraphics[draft=false]{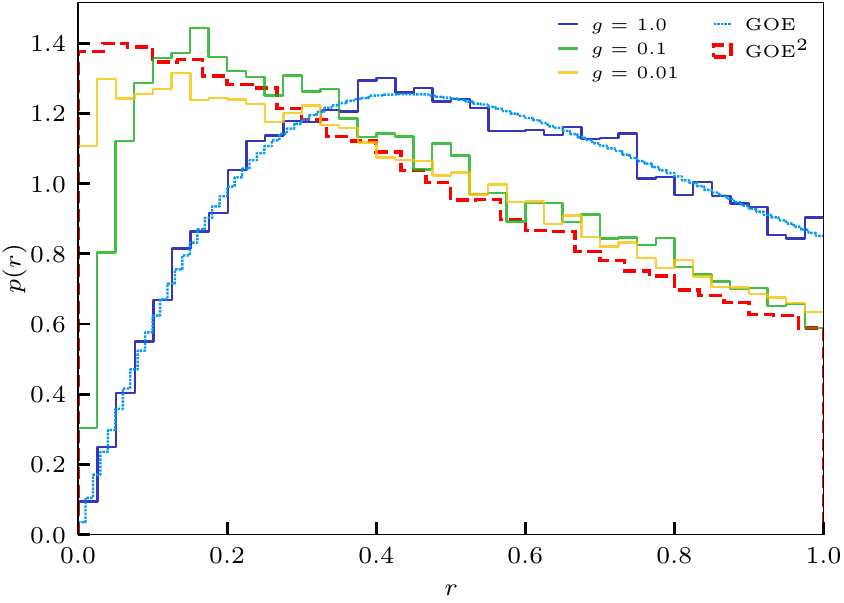}
    \caption{Spectral statistics of a single LIOM coupled to an $L_b=3$ ergodic region described by
    a GOE matrix. For a large enough coupling strength $g\gtrsim0.5$, the whole system follows the
GOE statistics, while for small coupling it reproduces the result expected in the completely
uncoupled case $g=0$, where the statistics corresponds to two ``folded'' GOE spectra.}
    \label{fig:gapsingle}
\end{figure}

\textbf{Single LIOM} ---
Here, we present additional numerical results for a single LIOM coupled to an ergodic grain of
size $L_b=3$. Our main focus is to understand how large the direct coupling $g$ has to be in order
for the full system to be ergodic. In Fig. \ref{fig:gapsingle}, we demonstrate that two limiting
cases are reached: For large enough coupling $g\gtrsim0.5$, the distribution of the adjacent gap
ratio parameter $r$ follows closely the GOE expectation, which means that the LIOM is well
thermalized by the ergodic grain. In the case of completely decoupled LIOM and ergodic grain, the
distribution is determined by a ``folding'' of two independent GOE spectra, denoted as GOE$^2$ in
Fig. \ref{fig:gapsingle}. Clearly for small coupling strengths, $g\lesssim 0.5$, the distributions
are quite close to this case and we conclude that if $g=0.1$, the LIOM is \emph{not} thermalized by
the ergodic region. This strenghtens our argument in the main text that for our longest chains and
$\alpha=0.75,0.8,0.85$, the last few spins are coupled so weakly to the ergodic region that they can
not be trivially thermalized.

\bibliography{goe_bubble}
\end{document}